\documentclass[aps,prl, twocolumn, superscriptaddress]{revtex4-1}
\usepackage{graphicx} %graphics handler
\usepackage{dcolumn}% Align table columns on decimal point
\usepackage{bm}% bold math
\usepackage{amssymb,amsmath}
\usepackage{epstopdf}
\usepackage{color}

\begin{document}

\title{Measurement of Filling-Factor-Dependent Magnetophonon Resonances in Graphene Using Raman Spectroscopy}
\author{Y.~Kim}
\author{J. M.~Poumirol}
\affiliation{National High Magnetic Field Laboratory, Tallahassee, FL 32310, USA}
\author{A.~Lombardo}
\affiliation{Cambridge Graphene Centre, University of Cambridge, 9 JJ Thomson Avenue, Cambridge, CB3 0FA, UK}
\author{N.~G.~Kalugin}
\affiliation{Department of Materials and Metallurgical Engineering, New Mexico Tech, Socorro, NM 87801, USA}
\author{T. Georgiou}
\author{Y.~J.~Kim}
\author{K.~ S.~Novoselov}
\affiliation{School of Physics \& Astronomy, University of Manchester, Oxford Road, Manchester M13 9PL, UK}
\author{A.~C.~Ferrari}
\affiliation{Cambridge Graphene Centre, University of Cambridge, 9 JJ Thomson Avenue, Cambridge, CB3 0FA, UK}
\author{J.~Kono}
\affiliation{Department of Electrical and Computer Engineering and Department of Physics and Astronomy, Rice University, Houston, TX 77005, USA}
\author{O.~Kashuba}
\affiliation{Institute for Theoretical Physics A, RWTH Aachen, D-52074 Aachen, Germany}
\author{V.~I.~Fal'ko}
\affiliation{Department of Physics, Lancaster University, Lancaster, LA1 4YB, UK}
\affiliation{DPMC, Universit\'e de Gen\`eve, CH-1211 Gen\`eve 4, Switzerland}
\author{D.~Smirnov}
\email{smirnov@magnet.fsu.edu}
\affiliation{National High Magnetic Field Laboratory, Tallahassee, FL 32310, USA}
\date{\today}
\begin{abstract}
We perform polarization-resolved Raman spectroscopy on graphene in magnetic fields up to 45T. This reveals a filling-factor-dependent, multicomponent anticrossing structure of the Raman $G$ peak, resulting from magnetophonon resonances between magnetoexcitons and $E_{2g}$ phonons. This is explained with a model of Raman scattering taking into account the effects of spatially inhomogeneous carrier densities and strain. Random fluctuations of strain-induced pseudomagnetic fields lead to increased scattering intensity inside the anticrossing gap, consistent with the experiment.
\end{abstract}


\maketitle
Magnetophonon resonances (MPRs) are observed in semiconductors when the energy of an optical phonon coincides with the inter-Landau level (LL) separation \cite{GurevichFirsov61}. This give important information on electron-phonon interactions, especially in two dimensional systems \cite{FirsovGurevich91}. Electron-phonon coupling (EPC) in graphene and graphite has been investigated for several years \cite{PiscanecetAl04PRL,FerrarietAl06PRL,Ando06JPSJ2,Ferrari07ssc,Basko09}. The zone-centre, doubly degenerate $E_{2g}$ phonon, strongly interacts with electrons, resulting in renormalization of phonon frequencies and line broadening \cite{PisanaetAlNatMat07,Yan07, DasetAl08NatureNano, DasetAl09PRB,YanPRB09, Lazzeri06}. These are tunable by electric and magnetic fields, through Fermi-energy shifts and Landau quantization. The Raman $G$ peak in graphene is predicted to exhibit anti-crossings when the $E_{2g}$ phonon energy matches the separation of two LLs \cite{AndoSLG04,Goerb07,AndoBLG04}. This MPR effect can be described as a resonant mixing of electronic and lattice excitations into a combined mode, leading to a splitting proportional to the EPC \cite{Goerb07}. It was observed in magneto-Raman scattering on single layer graphene (SLG) on the surface of graphite \cite{Yan10,Faug11}, non-Bernal stacked multilayer graphene on SiC \cite{Faug09}, and  bulk graphite \cite{Kim12}.

Here, we report a polarization-resolved Raman spectroscopy study of MPRs in SLG, demonstrating a strong dependence of the MPR line shape on the Raman polarization and carrier density. This is explained as a manifestation of MPRs between electronic magneto-excitons and circularly polarized optical phonons\cite{Goerb07}, combined with the effect of inhomogeneous carrier densities and strain.

The energy spectrum of SLG in a perpendicular magnetic field $B$ consists of discrete fourfold (spin and valley) degenerate Landau Levels (LL) with energies $E_n =\mathrm{sign}(n)\sqrt{2|n|}\hbar\tilde{c}/l_B$, where $n=...,-2, -1, 0, 1, 2, ..$ is the index of LLs in the conduction (n$>$0) and valence (n$<$0) bands, $n=0$ being exactly at the Dirac point (see Ref.\onlinecite{KNetoRMP09} and references therein). The band velocity, $\tilde{c}$, is the slope of the Dirac cone at zero $B$, and  $l_B=\sqrt{\hbar/eB}$ is the magnetic length. The LL occupancy is characterized by the filling factor $\nu=2\pi l_B^2 \rho_s$, where $\rho_s$ is the carrier density. Fully filled LLs with $n=0, 1, ...$ have $\nu =2, 6, ...$\cite{KNetoRMP09}. MPR in graphene occurs when the energy of the interband inter-LL transitions, $-n \to n\pm1$, matches that of the $E_{2g}$ phonons, resulting in strongly coupled electron-phonon modes, in which the electronic magneto-exciton is an anti-symmetric superposition of inter-LL excitations in each of the two SLG valleys, $K$ and $K'$, while this gives symmetric superposition active in far-infrared absorption\cite{Goerb07,Kashuba09}. This effect is strongest for the $E_{2g}$ MPR with inter-Landau-level transitions $-1 \to 0$  and $0 \to 1$, at $B^0_{MPR}\approx$25-30T.
Since the symmetry of such electronic excitations allows them to couple to the Raman-active $E_{2g}$ phonons \cite{Kashuba09},MPRs manifest through a fine structure which develops in the Raman $G$ peak in the vicinity of the MPR condition, $\Omega_{\Gamma}=[\sqrt{|n|}+\sqrt{|n|+1}]\sqrt{2}\hbar\tilde{c}/\l_B$ at $B^n_{MPR}$.
\begin{figure}
\centerline{\includegraphics[width=90mm]{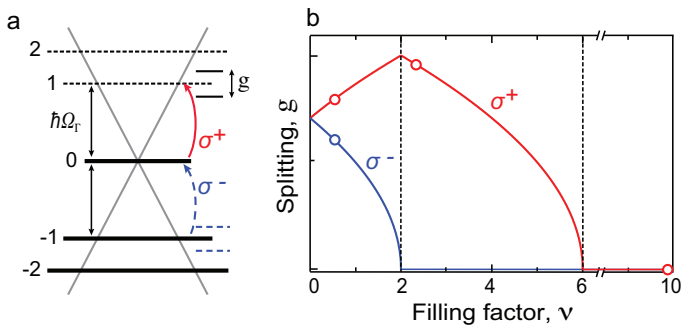}}
\caption{(a) (grey lines) SLG band structure at $B=0$.  Mode splitting ensues in the vicinity of MPR between $0 \to 1$ electron-hole excitations and $E_{2g}$ phonon, at $B^0_{MPR}\approx$25-30T. (b) Calculated \cite{Goerb07} mode splitting $g$ as a function of filling factor, $\nu$. Open circles indicate the filling factors probed in our experiment}
\label{Fig1}
\end{figure}

A specific feature of MPR in SLG, is the $\nu$ dependence of the anticrossing fine structure of coupled electron and phonon modes (which can be tuned externally by varying the carrier density), as well as a unique possibility to resolve the MPR of circularly polarized phonons\cite{Goerb07} (Fig.\ref{Fig1}). The polarization properties of Raman scattering involving magnetoexcitons are such that the incoming and outgoing photons have opposite circular polarizations, with angular momentum transfer $\pm2\hbar$, of which $\pm3\hbar$ is coherently transferred to the lattice, while the remaining angular momentum transfer $\mp1\hbar$ goes into electronic excitations\cite{Kashuba09}. In the following, we will refer to this angular momentum transfer as a $\sigma^{\mp}$-polarized transition between the electronic LLs, with opposite signs for $-n \to n\pm1$ inter-LL excitons: $\sigma^+$ for $-n \to n+1$ and $\sigma^-$ for $-n-1 \to n$ valence-to-conduction band transitions. Blocking such transitions by filling (depleting) the final states or depleting (filling) the initial states with electrons would suppress (promote) coupling between magneto-excitons and phonons. Since the $\sigma^+$ and $\sigma^-$ magneto-excitons are based on different LL pairs, changing carrier density in SLG would affect differently the size of the anticrossing in the MPR of $\sigma^{\mp}$ circularly polarized phonons, as illustrated in Fig.1. At $\nu =0$, corresponding exactly to a half-filled $n=0$ LL, the coupling strength of $\sigma^+$- and $\sigma^-$-polarized modes is equal, causing the $G$ peak to split equally for $\sigma^+$ and $\sigma^-$ phonon polarizations\cite{Goerb07}. For $0<\nu<2$, corresponding to a more than half-filled $n=0$ LL, the $-1 \to 0$ transition becomes partly blocked, while the $0 \to +1$ transition is promoted, giving rise to different splittings in the fine structure of $\sigma^+$-polarized and $\sigma^-$ polarized modes. For $2<\nu<6$, $n=0$ LL is full, leaving no space for MPR with the $\sigma^-$ phonon, whereas $n=+1$ LL is only partly filled, resulting in a MPR-induced fine structure in the $\sigma^+$ phonon lineshape (with maximum splitting at $\nu=2$). Thus, the EPC at the MPR resolves energies of the $\sigma^{\mp}$-polarized lattice excitations\cite{Goerb07,Kashuba09}. Note that, in order to probe the direct $B$ influence on the motion of lattice nuclei would require unattainably strong B$\sim$10$^4$T or higher, as can be estimated from the carbon-atom-to-electron mass ratio. Finally, at high $\nu>6$, both $-1 \to 0$ and $0 \to 1$ transitions are blocked, completely suppressing the MPR-induced fine structure for both modes.

We measure a SLG, CVD-grown on copper\cite{Li_s_2009,bae}, and transferred onto a Si/SiO$_2$ substrate\cite{Bonag}, following the procedures discussed in Supplemental Material\cite{Suppl}. The carrier density is estimated by combining the intensity and area ratio of the Raman $G$ and $2D$ peaks, I(2D)/I(G) and A(2D)/A(G), with the position and Full Width at Half Maximum of these peaks, Pos(G), Pos(2D), FWHM(G), FWHM(2D)\cite{DasetAl08NatureNano,PisanaetAlNatMat07,Basko09}. Under ambient laboratory conditions, the SLG sample is initially p-doped with a carrier concentration$\sim$5$\times$10$^{12}$cm$^{-2}$. By adjusting  annealing parameters (typically 8 hours in 90$\%$/10$\%$ Ar/H$_2$ atmosphere at temperatures up to 220$^{\circ}$C) and degassing in $<10^{-4}$mbar vacuum, the sample can be made n-type, with an electron density$\sim2\times10^{12}$cm$^{-2}$. Exposing the sample to low-pressure ($\sim$1mBar or less) $N_2$ atmosphere reduces the carrier density to$\sim0.4\times10^{12}$cm$^{-2}$. The exposure to ambient pressure $N_2$ gas or air results in hole doping, restoring p-type doping with carrier densities$\sim$5.5$\times$10$^{12}$cm$^{-2}$. If the sample is left in ambient air for a long period of time, it continues to experience p-doping reaching a carrier concentration$\geq$10$^{13}$cm$^{-2}$ in several weeks. However, this doping is inhomogeneous, with a 10 to 20\% variation, as shown by Raman mapping discussed in the Supplemental Material\cite{Suppl}.

$B$-dependent Raman measurements are performed in a quasibackscattering configuration at 300K for $B$ up to 45T using a high-field magneto-optical insert, as for Ref.\onlinecite{Kim12}. The combination of a linear polarizer and a $\lambda /4$ plate, circularly polarizes the incident and scattered light. The excitation beam (532nm) is focused to a spot$\lesssim$10$\mu$m, with a power$\sim$4mW. The circularly polarized $\sigma^\pm$ excitations, probed using the $\sigma^\pm/\sigma^\mp$ configurations of out/in polarisations, are achieved by reversing the $B$ polarity. In this notation, the first (second) symbol defines the polarization of the excitation (scattered) light. The polarization efficiency, i.e. the ratio of light transmitted through two circular polarizers with parallel or crossed helicities, is$\sim$90\% for the incident light, while it is$\sim$75\% for the scattered light in the spectral range of interest$\sim$1450-1700cm$^{-1}$, as discussed in Supplemental Material\cite{Suppl}. The spectral resolution is$\sim$1.9cm$^{-1}$.
\begin{figure*}
\centerline{\includegraphics[width=170mm]{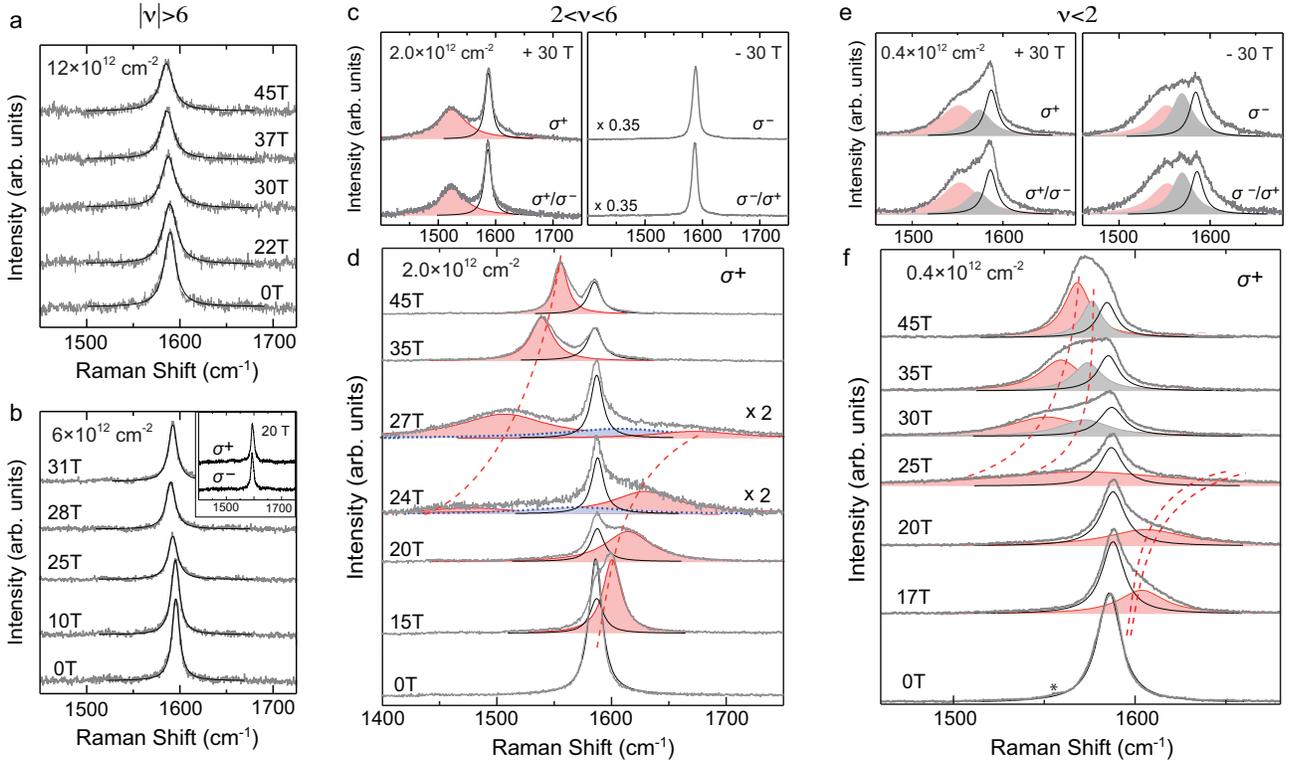}}
\caption{(a), (b) Unpolarized magneto-Raman spectra at carrier density corresponding to $\nu\gtrsim6$. The inset compares 20T $\sigma^+$- and $\sigma^-$ circular polarized spectra. (c-f) Circular-polarized magneto-Raman spectra at (c,d) $2<\nu<6$ and (e,f) $\nu<2$ for $B>15$T. (f) The star symbol points to a $B$-independent peak caused by parasitic scattering in fibers.This artifact, shown in the $B$=0 spectrum, is subtracted from all other spectra. The dashed lines in (d,f) are guides to the eye for the $B$ dependence of electron-phonon coupled modes at the $(0 \to 1)$  MPR anticrossing. The 
%blue 
dotted curves in (d) are additional spectral components observed in the close vicinity of the intersection between a $(0 \to 1)$ magneto-exciton and $E_{2g}$ phonon energy}
\label{Fig2}
\end{figure*}

\begin{figure}
\centerline{\includegraphics[width=90mm]{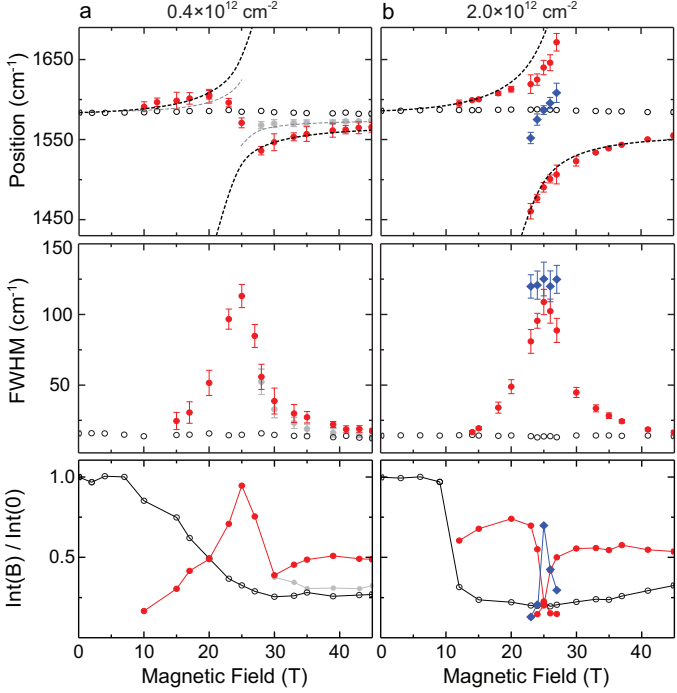}}
\caption{Peak position, FWHM and normalized intensity of the $G$ peak Lorenzian components as a function of B. Open black circles represent the central $E_{2g}$  phonon line component. Filled red and grey circles are the electron-phonon coupled modes. Dashed lines plot Eqs.\ref{eq:MPR_Esplitting}. Blue squares show additional components observed close to the resonance $B^0_{MPR}\approx25T$}
\label{Fig3}
\end{figure}

Figures \ref{Fig2} and \ref{Fig3} show the polarization and $B$ dependence of the $G$ peak at different carrier densities. The Raman spectra can be classified in three categories, as a function of $\nu$. At high carrier density, corresponding to $\left| \nu \right| \ge6$, the $G$ peak does not reveal MPR-induced splitting nor any polarization dependence, Figs.\ref{Fig2}a,b. As the filling factor decreases ($\nu<6$), a significant change in the spectra is observed. Figs.\ref{Fig2}c,d plot spectra measured on the sample annealed, degassed and kept in$<10^{-4}$mbar vacuum. The $G$ peak exhibits a strong, anticrossing-like splitting reaching$\sim150$cm$^{-1}$ ($\sim20$meV) for $B$=25T. The electron-phonon coupled modes appear only in the $\sigma^+$ and $\sigma^+/\sigma^-$ geometry, while the $G$ peak neither splits nor shifts in $\sigma^-$ and $\sigma^-/\sigma^+$ polarizations. This indicates n-type doping with estimated carrier density$\sim2.0\times 10^{12}$cm$^{-2}$. The observed behavior reveals the MPR at $2<\nu<6$. Indeed, the condition $\nu=6$ is met at $\sim B$=10T, with a coupling of $0 \to 1$ magnetoexcitons and E$_{2g}$ phonons at higher B. The MPR polarization selection rules is determined by the helicity of the incoming light only. As the carrier density further decreases, so that $\nu<2$, the G peak splitting changes (Figs.\ref{Fig2}e,f). In contrast to the $2<\nu<6$ case, the coupled modes now appear in both $\sigma^+$ and $\sigma^-$ polarizations. The spectra at $B\ge30$T reveal that the coupled-mode consists of two peaks, resolved in the cross-polarized measurements shown in Figs.\ref{Fig2}c,d (see also the second derivative data in Supplemental Materials\cite{Suppl}). The relative intensities of the two MPR spectral components depend on the incoming light helicity. At 20T,e.g., the expected separation of the two MPR components is$\sim$20cm$^{-1}$, while their linewidths estimated from the 20T, $2<\nu<6$ spectra, are$\sim$50cm$^{-1}$. We assign these two components to the $\sigma^\pm$ polarized modes originating from the coupling of $E_{2g}$ phonons with $0 \to 1$ and $-1 \to 0$ magneto-excitons. The fact that $\sigma^\pm$ polarized modes do not completely disappear in the measurements with opposite helicity as due to the imperfect polarization selectivity of our setup, combined with sample inhomogeneity, as discussed below.

For the strongest anticrossing, at $B^0_{MRP}$, the observed behavior can be described using the resonance approximation formula\cite{Goerb07} for pairs of strongly coupled $\Omega^{\sigma}_{\pm}$ circularly polarized modes:
\begin{gather}
\label{eq:MPR_Esplitting}
\Omega^{\sigma}_{\pm} =\frac{\Omega_{\Gamma}+\Omega_{0}}{2}\pm\sqrt{\left(\frac{\Omega_{\Gamma}-\Omega_{0}}{2}\right)^ 2+\frac{\mathnormal{g_{\sigma}}^2}{\hbar^{2}}}, \\
g_{\sigma^{+}}(2<\nu<6) = g_{0} \frac{a}{l_{B}} \sqrt{6-\nu},
\qquad
g_{\sigma^{-}}(2<\nu<6) =0, \nonumber \\
g_{\sigma^{\pm}}(0<\nu<2) =  g_{0} \frac{a}{l_{B}} \sqrt{2\pm\nu},
\qquad
g_0 =  \sqrt{\frac{3}{2}}\sqrt{\frac{\lambda_{\Gamma}}{4\pi}}\gamma_{0}. \nonumber
\end{gather}
Here, $\Omega^{\sigma}_\pm$ is the frequency of the upper (+) or lower (-) coupled mode, $\Omega_{\Gamma}\sim1581$cm$^{-1}$ is the bare $E_{2g}$ phonon energy, $\Omega_{0}=\sqrt{2}\hbar\tilde{c}/\l_B$, $a$ ($=2.46$\AA) is the graphene lattice constant. $g_0$ describes the EPC at zero $B$ in terms of the nearest neighbor hopping integral $\gamma_{0}$, and the dimensionless coupling constant $\lambda_{\Gamma}$\cite{Basko09,Lazzeri06}.

The set of equations \eqref{eq:MPR_Esplitting} describe the energies of the two branches of mixed phonon and magneto-exciton modes for an idealized non-strained, homogeneously doped SLG, also neglecting broadening of the bare excitations due to inelastic and elastic scattering. The detailed description of MPR and its manifestation in the Raman spectrum of realistic samples should include broadening of the optical phonon, $\gamma_\Gamma$, and of the magneto-exciton, $\gamma_{e}$, as well as possible inhomogeneity of doping and fluctuations of local strain. In a sample with a spatially varying carrier density, $g^{\sigma^{+}}$ and $g^{\sigma^{-}}$ become coordinate-dependent, which would lead to inhomogeneous broadening of the MPR fine structure. Inhomogeneous strain in SLG generates a pseudo-magnetic field, $B_{str}$ with opposite sign in the $K$ and $K'$ valleys\cite{Iord85,Ando06JPSJ2,Morp06}, giving a total field $B_K=B+B_{str}$  at $K$ and $B_{K'}=B-B_{str}$ at $K'$.
In strained graphene nanobubbles, it was reported that $B_{str}$ can reach 200-400T\cite{Levy10}. In nominally unstrained SLG, the residual strain caused by nm-scale height variations and defects can induce random $B_{str}>$10T\cite{Yeh}. Locally, this splits a simple anticrossing into mixing and splitting of each $\sigma^\pm$-polarized phonon and two magneto-excitons, now distinguished by different LL energies in the two valleys, leading to an even more complicated fine structure.
Also, the quantum efficiency (i.e. the probability of a scattering event per incoming photon) of $0 \to 1$ and $-1 \to 0$ magneto-excitons in Raman scattering\cite{Kashuba09} is much lower than the quantum efficiency of the $\Gamma$ optical phonon, manifested in the $G$ peak area, A(G)\cite{BaskoNJP09}.

All these factors can be taken into account in the formula for the spectral weight of Raman scattering at MPR derived using the random phase approximation\cite{KashubaNJP12}:
\begin{equation}
I^{\sigma^{\pm}}(\omega) = \frac{A(G)}{2\pi}
\Biggl<\mathrm{Im}
\frac{1}{\omega-\Omega_{G}-i\gamma_{\Gamma} \mp \frac{\lambda_{\Gamma}}{4\pi}\sum_{\alpha}
\frac{\Omega_{0,\alpha}^{2}}{\omega-\Omega_{0,\alpha}-i\gamma_{e}}}
\Biggr>,
\label{eq:MPR}
\end{equation}
where angle brackets stand for the averaging over the spatial fluctuations of a strain-induced $B_{str}$ with zero average, and over fluctuations of electron density.
These together determine the local values of the inter-LL separation, $\Omega_{0,\alpha}=\sqrt{2}\hbar\tilde{c}/l_{B_\alpha}$, and the local filling factors of the $n=0$ and $n=\pm1$ LLs in each of the two valleys $\alpha=K, K'$. Eq.\eqref{eq:MPR_Esplitting} describing energies of coupled modes can also be obtained from Eq.\eqref{eq:MPR} neglecting strain and carrier density fluctuations.

To measure the EPC, we fit Eq.\eqref{eq:MPR_Esplitting} to the data of Fig.\ref{Fig3}. The high-field anticrossing branches ($\Omega_{-}$) are most accurately resolved and therefore most suitable for fitting. The energy of the unperturbed electronic transition $\Omega_0$ is calculated using $\tilde{c}=1.08\times10^6$m/s, which corresponds to the MPR at 25T. For low carrier density ($0.4\times 10^{12}$cm$^{-2}$), we obtain $g_0^{\sigma^+}= 17$cm$^{-1}/$T$^{1/2}$ and $g_0^{\sigma^-}=12$cm$^{-1}/$T$^{1/2}$. For intermediate carrier density ($2.0\times 10^{12}$cm$^{-2}$), fitting yields $g_0^{\sigma^+}=21$cm$^{-1}/$T$^{1/2}$. Plugging the fitted $g^{\sigma^{\pm}}$ into Eq.\eqref{eq:MPR_Esplitting}, we find $\nu$ at the resonance B, and extract carrier densities$\sim0.4\times 10^{12}$ and$\sim1.6\times 10^{12}$cm$^{-2}$ for the low and intermediate density samples, respectively, in excellent agreement with those derived from the Raman spectra. Taken together, these results allow us to deduce the $\lambda_{\Gamma}$. Notably, the values we obtain for the low and intermediate density samples differ by less than 5\%. We get $\lambda_{\Gamma}\approx 0.028$, in remarkable agreement with that measured from FWHM(G) in undoped samples at B=0, and predicted by density functional theory\cite{PiscanecetAl04PRL,Lazzeri06,Basko09}.
\begin{figure}
\centerline{\includegraphics[width=90mm]{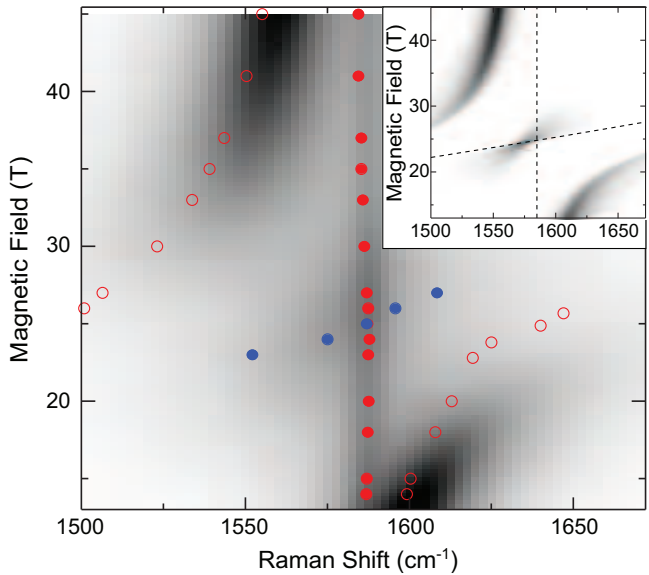}}
\caption{MPR in SLG with $\langle\rho_{s}\rangle=1.6\times 10^{12}$cm$^{-2}$ in the presence of inhomogeneous strain, $\langle B_{str}^2\rangle^{1/2}=6$T, calculated using Eq.\eqref{eq:MPR} with $\Omega_{\Gamma}\sim1581$cm$^{-1}$, $\gamma_{\Gamma}\sim12$cm$^{-1}$, $\gamma_{e}=50$cm$^{{-1}}$, and $\langle\delta\rho_s^{2}\rangle^{1/2}=3\times10^{11}$cm$^{-2}$. Symbols are experimental results. The inset shows the calculated Raman scattering intensity, neglecting LL broadening, depolarisation, inhomogeneity, but with strain-induced inhomogeneous pseudomagnetic fields. Dashed lines indicate energies of unperturbed E$_{2g}$ phonons and (0,1) magneto-excitons}
\label{Fig4}
\end{figure}

Finally, we discuss on the observation of Raman scattering in the middle of the MPR anticrossing gap. The spectra for the intermediate carrier density show an additional component, indicated 
%in blue 
by dotted curves in Fig.\ref{Fig2}d. This is detected over a narrow $B$ range, between 23 and 27T, reaching maximum intensity at 25T. Though we are not able to spectrally resolve this component at lower carrier density, a similar sharp increase of intensity is observed in the vicinity of the MPR at 25T (Fig.\ref{Fig3}a, bottom panel). Analysis of Eq.\eqref{eq:MPR} reveals that a finite Raman scattering inside the anticrossing gap is a specific signature of MPR in strained SLG. Fig.\ref{Fig4} plots the scattering intensity calculated for intermediate carrier density. To model the experiment, we assume density fluctuations described by a normal distribution with a $\langle\rho_s\rangle\sim1.6\times 10^{12}$cm$^{-2}$ average and 20$\%$ standard deviation, imperfect light polarization leading to a 75$\%$/25$\%$  mixture of $\sigma^+$ and $\sigma^-$ spectra, LL and E$_{2g}$ phonon broadening of $\gamma_{e}$=50cm$^{-1}$ and $\gamma_{\Gamma}$=12cm$^{-1}$ and, most importantly, inhomogeneous strain, $\langle B_{str}^2 \rangle^{1/2}=6$T. Away from the $\Omega_0=\Omega_{\Gamma}$ MPR intersection point, the scattering intensity follows the expected anticrossing MPR behavior, while the increased scattering in the middle of the anticrossing gap is due to the overall effect of inhomogeneous strain. This is further illustrated in the inset in Fig.\ref{Fig4}, with a scattering intensity map calculated for the ideal case of uniformly doped SLG ($\delta \rho_s=0$) exposed to the same random $B_{str}$, neglecting broadening ($\gamma_{\Gamma}=\gamma_{e}$=0) and depolarization.

In summary, we used polarization-resolved high-field magneto-Raman spectroscopy to investigate magneto-phonon resonances in graphene. By varying the filling factor, we identified different types of $G$ peak magnetic-field dependencies, providing a comprehensive experimental evidence of MPR on circularly polarized phonons. We also detected an unexpected increase of Raman intensity in the middle of the MPR anticrossing gap and assigned it to mixing and splitting of electron-phonon coupled modes caused by fluctuations of strain-induced pseudo-magnetic fields.

We acknowledge funding from NHMFL UCGP-5068, DOE/BES DE-FG02-07ER46451, DOE/BES DE-FG02-06ER46308, the Robert A.~Welch Foundation C-1509, EPSRC grants GR/S97613/01, EP/E500935/1, EP/K01711X/1, EP/K017144/1, EP/G042357/1, ERC grant NANOPOTS, EU grants GENIUS and RODIN, the Leverhulme Trust,  Royal Society , NSF ECCS 0925988, ERC AdG 'Graphene and Beyond', the NSF Cooperative Agreement No.~DMR-0654118, the State of Florida, the DOE.

\end{document}